\documentclass[%
 reprint,%
 amssymb, amsmath,%
 aip,cha,%
]{revtex4-1}

\usepackage{graphicx}% Include figure files
\usepackage{docs}%
\usepackage{bm}%
\usepackage[colorlinks=true,linkcolor=blue]{hyperref}%
\expandafter\ifx\csname package@font\endcsname\relax\else
 \expandafter\expandafter
 \expandafter\usepackage
 \expandafter\expandafter
 \expandafter{\csname package@font\endcsname}%
\fi
\begin{document}

\title{Measuring mode indices of a partially coherent vortex beam with HBT type experiment}%

\author{Ruifeng Liu}
\author{ Feiran Wang}
\author{Dongxu Chen}
\author{Yunlong Wang} 
\author{Yu Zhou} 
\author{Hong Gao} 
\author{Pei Zhang}
\email{zhangpei@mail.ustc.edu.cn}
\author{Fuli Li}

\affiliation{Key Laboratory of Quantum Information and Quantum Optoelectronic Devices, Shaanxi Province, Xi'an
Jiaotong University, Xi'an 710049, China}%

\date{\today}% It is always \today, today,
%\revised{August 2010}%

\begin{abstract}
It is known that the cross-correlation function (CCF) of a partially coherent vortex (PCV) beam  shows a robust link with the radial and azimuthal mode indices. However, the previous proposals are difficult to measure the CCF in practical system, especially in the case of astronomical objects. In this letter, we demonstrate experimentally that the Hanbury Brown and Twiss effect can be used to measure the mode indices of the original vortex beam and investigate the relationship between the spatial coherent width and the characterization of CCF of a PCV beam. The technique we exploit is quite efficient and robust, and it may be useful in the field of free space communication and astronomy which are related to the photon's orbital angular momentum.
\end{abstract}

\maketitle

%\tableofcontents

In 1992, L. Allen \textit{et al.} \cite{Allen} pointed out that beam with spiral phase distribution of $\exp(il\varphi)$ carries an orbital angular momentum (OAM) of $l\hbar$, where $l$ is an integer which denotes the azimuthal mode index (topological charge), and $\varphi$ is the azimuthal coordinate. Beam with this kind of phase structure is also called vortex beam. Vortex beam has been widely studied in the last two decades and found a lot of applications, such as optical tweezers \cite{Grier}, spiral phase contrast microscopy\cite{Jesacher}. Since the values of $l$ is theoretically unlimited, OAM states construct an infinite dimensional Hilbert space, and it exhibits great potential for applications in the field of quantum information process\cite{ENagali09,Nagali09}, free-space information transfer and communications\cite{Gibson,Vallone,Krenn}. Despite the extensive applications, determining the topological charge $l$ of OAM state remains an intriguing problem, and a lot of methods has been proposed. Such as Mach-Zehnder interferometer\cite{Leach}, diffraction pattern with specific masks\cite{Berkhout08,Hickmann,Liu13,Fu15}, image reformatting\cite{Berkhout}, intensity analysis\cite{Long,Xu14}.

Recently, more and more researches have focused on the partially coherent vortex (PCV) beam \cite{Palacios04,Dijk09,Kumar10,Jesus-Silva,Yang12,Zhao12,Yang13,Reddy13,Feng14,RKSingh14,Chen14,Alves15,Salla15,RKSingh15,Ding12}. In 2004, Palacios \textit{et al.}  firstly verified that a robust phase singularity exists in the spatial coherence function when a vortex is presented \cite{Palacios04} in the original beam. Meanwhile, in Refs. \cite{Yang12,Zhao12,Yang13}, the authors revealed the linkage between the mode indices and the cross correlation function (CCF) of a PCV beam, and they also discussed the spatial coherence on determining the mode indices of a PCV beam. 

Also, there has been interest in the use of spatial modes, such beams carrying OAM, as an additional degree of freedom to increase the available information bandwidth for free-space communication. Because a large Hilbert space is helpful to improve the security of cryptographic keys transmitted with a quantum key distribution system\cite{Bourennane01}. Since the important application of free space communication,  the effects of propagation through random aberrations (atmospheric turbulence) on coherence for single-photon communication systems based on orbital angular momentum states are also investigated \cite{Paterson05,Rodenburg12,Rodenburg14,Goyal14}. However, determining the mode indices of a PCV beam is still difficult. In Refs. \cite{Palacios04,Yang13}, the authors experimentally used a wavefront folding interferometer to measure the mode indices of a PCV beam. When the size of PCV beam is large compared to the measuring instrument, their scheme is difficult be used to measure the cross-correlation function. Meanwhile, the interference pattern in their results is complex and the visibility is low, which is difficult to characterize the mode indices from the experimental results. In this letter, we experimentally prove that the mode indices of a PCV beam can be measured efficiently through a Hanbury Brown and Twiss (HBT) type experiment which was firstly used in the field of astronomy \cite{Brown177,Brown178}.

An incoherent or partially coherent source both tend to emit photons together (bunching) with enhanced photon number fluctuations relative to classical expectations (super-Poisson statistics). This kind of bunching effect was firstly measured in the experiment of observing the second-order temporal and spatial intensity correlations of star light by Hanbury Brown and Twiss \cite{Brown177,Brown178}. It was believed that the nontrivial HBT correlation is caused by the measured intensity fluctuations of the thermal light. In the HBT experiment, the second-order correlation function is expressed as
\begin{eqnarray}
{G^{(2)}}({{\vec r}_1},{{\vec r}_2}) &=& \left\langle {I({{\vec r}_1})I({{\vec r}_2})} \right\rangle  \notag\\
&=& \left\langle {I({{\vec r}_1})} \right\rangle \left\langle {I({{\vec r}_2})} \right\rangle  + {\left| {\Gamma ({{\vec r}_1},{{\vec r}_2})} \right|^2},
\label{00}
\end{eqnarray}
where $ \langle ... \rangle $ denotes the ensemble average, $ I({{\vec r}_1}) $ and $ I({{\vec r}_2}) $ are the intensity at spatial point $ {{\vec r}_1} $ and $ {{\vec r}_2} $, respectively. $ \Gamma ({{\vec r}_1},{{\vec r}_2}) $ is the mutual coherence function (MCF) of the light source. Since the second-order correlation function in Eq. (\ref{00}) can be easily obtained by the coincidence measurement of two detectors, we may get the mode indices of PCV beam from the measured magnitude square of MCF, $ \left| {\Gamma ({{\vec r}_1},{{\vec r}_2})} \right|^2 $.

Laguerre-Gaussian (LG) modes are circularly symmetric and structurally stable solutions of the paraxial wave equation. The electromagnetic field amplitude of LG modes in $ z=0 $ is given by
\begin{eqnarray}
u_p^l(r,\varphi ,0) \propto {(\frac{r}{{{w_0}}})^l}L_p^l(\frac{{2{r^2}}}{{w_0^2}}){e^{ - {{{r^2}} \mathord{\left/
 {\vphantom {{{r^2}} {w_0^2}}} \right.
 \kern-\nulldelimiterspace} {w_0^2}}}}{e^{il\varphi }},
\label{01}
\end{eqnarray}
where $ r $ and $ \varphi $ are the radial  and azimuthal coordinates, respectively. $ p $ is the radial mode index, and $ l $ is the azimuthal mode index which describes the phase structure. $ w_0 $ denotes the waist width and $ L_p^{\left| l \right|} (...)$ is the generalised Laguerre polynomial.

\begin{figure}[htb]
\centerline{\includegraphics[width=8cm]{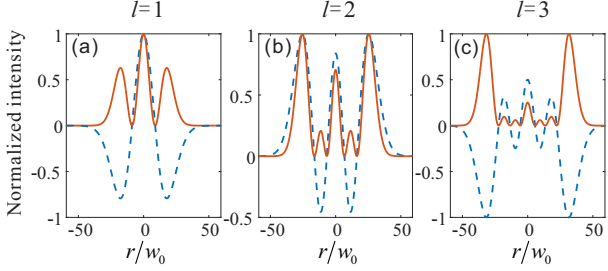}}
\caption{The CCF and magnitude square of CCF for the far-field partially coherent Laguerre-Gaussian modes. The blue-dash line and orange-solid line denote the $ \Gamma_{c} '({{\vec r'}},{-{\vec r'}})$ and $ \left| \Gamma_{c} '({{\vec r'}},{-{\vec r'}})\right|^2 $, respectively. Mode indices for $ p=0 $, and (a) $ l=1 $; (b) $ l=2 $; (c) $ l=3 $.}
\label{1d}
\end{figure}

A partially coherent LG beam can be generated by propagating the coherent LG beam through a rotating ground glass disk (GGD). It is well known that the statistical distribution \cite{Goodman} of the generated beam corresponds to a Gaussian-Schell correlator $ C({{\vec r}_1},{{\vec r}_2}) = \exp [ - {\left| {{{\vec r}_1} - {{\vec r}_2}} \right|^2}/\sigma _g^2] $, where $ \sigma _g $ represents the transverse spatial coherence width. Combining Eq. (\ref{01}) with $ C({{\vec r}_1},{{\vec r}_2}) $, the MCF of the partially coherent LG at source plane $ z=0 $ can be expressed as \cite{Yang13}
\begin{eqnarray}
\Gamma ({\vec r_1},{\vec r_2}) &\propto & C({\vec r_1},{\vec r_2}){(\frac{{{{\vec r}_1} \cdot {{\vec r}_2}}}{{w_0^2}})^l}L_p^{\left| l \right|}(\frac{{2\vec r_1^2}}{{w_0^2}})L_p^{\left| l \right|}(\frac{{2\vec r_2^2}}{{w_0^2}})
\notag\\
&& \times {e^{ - {{(\vec r_1^2 - \vec r_2^2)} \mathord{\left/
 {\vphantom {{(\vec r_1^2 - \vec r_2^2)} {w_0^2}}} \right.
 \kern-\nulldelimiterspace} {w_0^2}}}}{e^{il({\varphi _2} - {\varphi _1})}}.
\label{02}
\end{eqnarray}

After propagating a distance $ z $ in the far field, the MCF can be written as
\begin{eqnarray}
\Gamma '({{\vec r'}_1},{{\vec r'}_2}) &\propto & \frac{1}{{{\lambda ^2}{z^2}}}\int {\int {\int {\int {d{{\vec r}_1}d{{\vec r}_2}\Gamma ({{\vec r}_1},{{\vec r}_2})} } } } \notag\\
&& \times \exp [ - i\frac{{k }}{{ z}}({{\vec r'}_1} \cdot {{\vec r}_1} - {{\vec r'}_2} \cdot {{\vec r}_2})],
\label{03}
\end{eqnarray}
where $k $ is the wave number. As we known, the far field CCF, $ \Gamma_{c} '({{\vec r'}},{-{\vec r'}})$, shows a closed relationship with the topological charge of the original vortex beam, and it has been proved that the CCF will maintain the dislocation rings. The blue-dash line in Figs. \ref{1d} (a)-(c) show some simulation results of the far field one dimensional CCF for mode indices $ p=0$, $l=1,\;2,\;3 $ LG beams. In the simulation, the waist width to spatial coherence width ratio is set as $ w_0 /\sigma _{g}=0.5 $. It is straightforward to find that the number of dislocation rings in the $ \Gamma_{c} '({{\vec r'}},{-{\vec r'}})$ is equal to the value of azimuthal mode index $ l $ of the LG mode. Since the direct measurement of the CCF is troublesome, we will discuss and measure the $ \left| \Gamma_{c} '({{\vec r'}},{-{\vec r'}})\right|^2 $ in the following. As shown in Fig. \ref{1d}, the orange-solid lines represent the magnitude square of the CCF, $ \left| \Gamma_{c} '({{\vec r'}},{-{\vec r'}})\right|^2 $, which have the same number of dislocation rings as the original far field CCF. It means that the HBT effect of the partially coherent beam could be used to reveal the features of the CCF indirectly.

\begin{figure}[htb]
\centerline{\includegraphics[width=7cm]{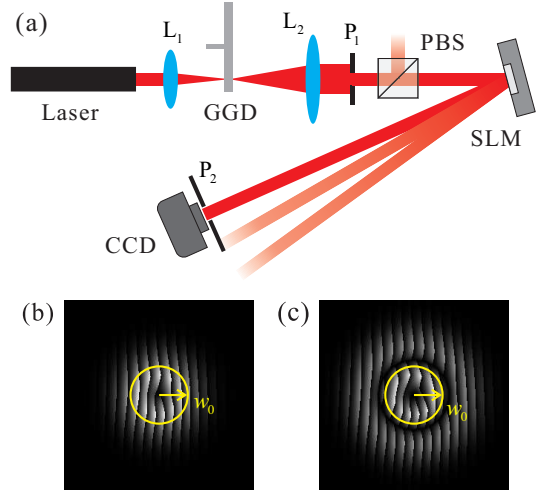}}
\caption{(a) Schematic of experimental setup to measure the mode indices of a partially coherent vortex beam. L$ _1 $ and L$ _2 $ are lenses. P$ _1 $ and P$ _2 $ are pinholes. PBS is polarization beam splitter and SLM is the phase spatial light modulate. CCD represents charge-coupled device camera. (b) and (c) are the the examples of holograms to generate LG beams with waist width $ w_0 $, and their mode indices are ($ p=0,\;l=1 $)  and ($ p=1,\;l=2 $), respectively. }
\label{setup}
\end{figure}

The experimental setup which uses the HBT effect to measure the mode indices of a PCV beam is shown in Fig. \ref{setup} (a). A coherent Gaussian beam is generated from the He-Ne laser with wavelength $ \lambda =633$ nm, then it is focused on a GGD by lens L$ _1 $. The angular velocity of the GGD is kept at $ \omega=\pi$ rad/s, and the spatial coherence width $ \sigma _{g} $ can be controlled by changing the distance between lens L$ _1 $ and GGD. Lens L$ _2 $ and pinhole P$ _1 $ are used to collimated the partially coherent beam after the GGD. Then the beam is filtered by a polarization beam splitter (PBS) to the horizontal polarization to match the requirement of phase spatial light modulator (SLM). Figs. \ref{setup} (b) and (c) display the example computer-generated hologram  gratings \cite{Bolduc13} which are imprint on  the phase SLM (Hamamatsu model X10468) to generate the partially coherent LG beam. The gratings correspond to mode indices ($ p=0,\;l=1 $)  and ($ p=1,\;l=2 $), respectively.  The waist width $ w_0 $ of partially coherent LG can also be handily controlled  by changing the gratings on computer, and the average full width at half maximum of the speckles is 0.34 mm. The first-order diffracted beam is selected out by a pinhole P$ _2 $, and the intensity pattern is recorded by a charge-coupled device (CCD). The CCD is a $ 1040\times1392 $ array of $6.45\times 6.45 \;\mu$m$^2$ pixels, and the measurement is made with an exposure time of $ 0.2$ ms to guarantee that the acquired images are temporally incoherent. The field on the plane of the CCD can be regarded as the partially coherent LG beam. Instead of using two point detectors for the coincidence measurement of the second-order correlation function in Eq. (\ref{00}), a CCD is used here to recorded the intensity pattern and then analyse the second-order correlation on computer \cite{rui04}.

\begin{figure}[htb]
\centerline{\includegraphics[width=6.5cm]{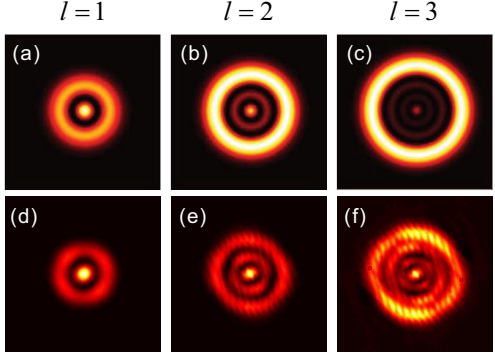}}
\caption{Magnitude square of cross correlation function for the far-field partially coherent Laguerre-Gaussian ($ p=0,\;l=1,2,3 $) modes. (a)-(c) and (d)-(f) are simulated and experimental results, respectively. The results correspond to waist width $ w_0 =0.5\sigma _{g} $.}
\label{result1}
\end{figure}

To acquire the $ \left| \Gamma_{c} '({{\vec r'}},{-{\vec r'}})\right|^2 $ of the partially coherent LG beam with the setup in Fig. \ref{setup} (a), we record $N=5000$ frames of images in each experiment. According to definition of intensity correlation in Eq. (\ref{00}), the recorded data for the magnitude square of CCF are processed as follows
%\begin{eqnarray}
\begin{align}
&{\left| {{\Gamma _c}(\vec r', - \vec r')} \right|^2}   \notag\\
&=\frac{1}{N}\sum\limits_{i = 1}^N {{I_i}(\vec r'){I_i}( - \vec r')}  - \frac{1}{{{N^2}}}\sum\limits_{i = 1}^N {{I_i}(\vec r')} \sum\limits_{i = 1}^N {{I_i}( - \vec r')}, 
\label{04}
\end{align}
%\end{eqnarray}
where $ {I_i}(\vec r) $ is the $ i $th frame of intensity distribution on the CCD plane and $ \vec r $ denotes the pixel coordinate of the CCD. Before recording the data for $ \left| \Gamma_{c} '({{\vec r'}},{-{\vec r'}})\right|^2 $, the CCD is placed beforehand on the plane of the SLM in Fig. \ref{setup} (a) to measure spatial coherent width $ \sigma _g $. 

\begin{figure}[htb]
\centerline{\includegraphics[width=8cm]{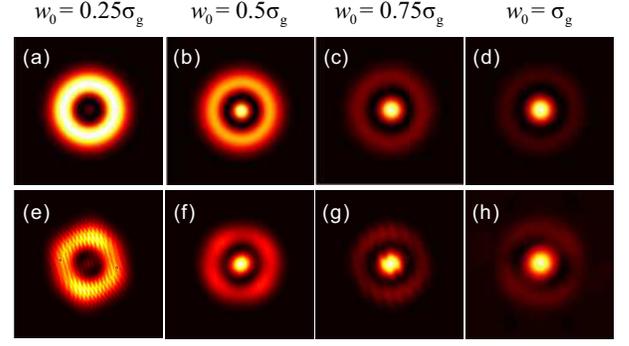}}
\caption{Magnitude square of the cross correlation function for the partially coherent Laguerre-Gaussian ($ p=0,\;l=1 $) mode with different waist width to spatial coherence width ratio. (a)-(d) and (e)-(h) are corresponding to simulated and experimental results, respectively.}
\label{result2}
\end{figure}

Fig. \ref{result1} shows the theoretically and experimentally obtained two dimensional $ \left| \Gamma_{c} '({{\vec r'}},{-{\vec r'}})\right|^2 $ of the far-field partially coherent LG beams with mode indices $ p=0,\;l=1,\;2,\;3 $ and waist width $ w_0 =0.5\sigma _{g} $. Here we see that the number of dislocation rings in $ \left| \Gamma_{c} '({{\vec r'}},{-{\vec r'}})\right|^2 $ matches with the azimuthal mode index $ l $ of the original LG beam. The results are consistent with the earlier theoretical predication in Ref. \cite{Yang12}, and it means that one can measure the azimuthal index of a PCV beam through the HBT type measurement.

Furthermore, we investigate the character of $ \left| \Gamma_{c} '({{\vec r'}},{-{\vec r'}})\right|^2 $ under different waist width to spatial coherence width ratios. Figs. \ref{result2} (a)-(d) and (e)-(f) show the simulated and experimental results with LG mode $ p=0,\;l=1 $, and they are corresponded to 4 typical waist width to spatial coherence width ratios. The experimental patterns agree well with the theoretical distributions. From Fig. \ref{result2} we can see that the bright spot in the center of the $ \left| \Gamma_{c} '({{\vec r'}},{-{\vec r'}})\right|^2 $ will gradually diminish in radius as the increase of the spatial coherent width, and eventually vanish in the situation of a coherent LG beam. On the contrary, the outside rings also will be too blurry to be invisible as the spatial coherence of the beam become quite low. The high and low coherence cases are shown in Figs. \ref{result2} (e) and (h), respectively. In both of these two situations, it might be difficult to count the number of dislocation rings for low contrast ratio.

\begin{figure}[htb]
\centerline{\includegraphics[width=7.5cm]{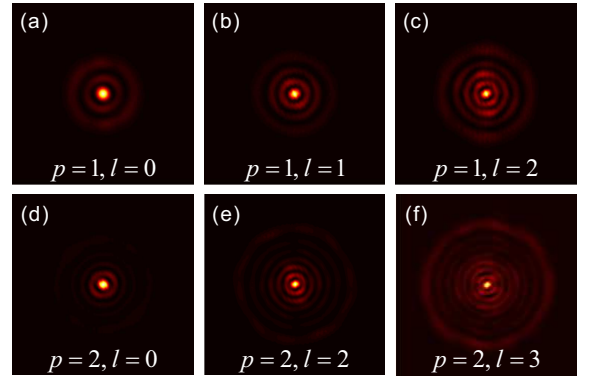}}
\caption{Experimental results of magnitude square of the cross correlation function for the higher-order partially coherent Laguerre-Gaussian modes. Their waist width to spatial coherence width ratios are $ w_0 /\sigma _{g}=0.5 $. To show the picture clearly, we reduce the contrast of the picture by setting the values of $ \left| \Gamma_{c} '({{\vec r'}},{-{\vec r'}})\right|^2 =0.3* \max \{ \left| \Gamma_{c} '({{\vec r'}},{-{\vec r'}})\right|^2  \}  $ where $ \left| \Gamma_{c} '({{\vec r'}},{-{\vec r'}})\right|^2 >0.3* \max \{ \left| \Gamma_{c} '({{\vec r'}},{-{\vec r'}})\right|^2  \}  $.}   
\label{result3}
\end{figure}

In Ref. \cite{Yang13}, the authors investigated the linkage between the number of dislocation rings $ \Omega $ in the far-field cross correlation function and the mode indices of a partially coherent LG beam, and they showed that $ \Omega=2p+\left| l \right| $. Their experiments demonstrated that the spatial correlation singularity still exist even for a non-vortex ($ l=0 $) partially coherent beam if the radial index $p$ is non-zero. Likewise, we experimentally prove the relationship $ \Omega=2p+\left| l \right| $ for some higher-order partially coherent LG beams with waist width to spatial coherence width ratios to be $ w_0 /\sigma _{g}=0.5 $, and the results are shown in Fig. \ref{result3} which have high visibility. Figs. \ref{result3} (a) and (d) show the $ \left| \Gamma_{c} '({{\vec r'}},{-{\vec r'}})\right|^2 $ of non-vortex partially coherent LG beam with radial index $ p=1 $ and $ p=2 $, respectively. It is obvious that the spatial correlation singularity still exists and the number of dislocation rings is $ 2p $. We also measured the $ \left| \Gamma_{c} '({{\vec r'}},{-{\vec r'}})\right|^2 $ of higher-order partially coherent LG modes which are shown in Figs. \ref{result3} (b), (c), (e), (f).  A desirable results in the experiment are achieved, and it is easy to count the number of dislocation rings as $ \Omega=2p+\left| l \right| $. 

In conclusion, we experimentally show the relationship between the number of dislocation rings of the magnitude square of the CCF and the radial and azimuthal mode indices ($ p,\;l $) of a PCV beam through a HBT type experiment, and our results prove that the number of dislocation rings is identical to $ 2p+\left| l \right| $. We also investigate the effect of spatial coherent width on determining the mode indices of the PCV beams. Our method offers a powerful tool to investigate the character of the MCF of a PCV beam and determine the mode indices of the PCV beam, which is efficient and robust in practical measurement system because only the HBT type setup is used to get the second-order intensity correlation. It has potential applications in the field of free space communication and astronomical research with the photon's OAM degree of freedom, especially in the situation with non-ignorable effect of atmospheric turbulence scattering medium.

\section*{Funding Information}

This work is supported by the Fundamental Research Funds for the Central Universities, Program for
Key Science and Technology Innovative Research Team of Shaanxi Province (2013KCT-05), and the
National Natural Science Foundation of China ( Grant Nos. 11374008, 11374238, 11374239 and 11534008).

\end{document}